\documentclass[preprint,showpacs,amsmath,amssymb]{revtex4}

\usepackage{graphicx,amsmath,amssymb,bm,multirow}
\usepackage{amsthm}
\usepackage{amsfonts}
\usepackage{dcolumn}
\usepackage{bm}
\usepackage[colorlinks,linkcolor=blue,anchorcolor=blue,citecolor=blue]{hyperref}
\usepackage[usenames,dvipsnames]{color}
\usepackage{CJK}
\usepackage{ulem}

\begin{document}

\title{Effect of pairing correlations on nuclear low-energy structure: BCS and general Bogoliubov transformation}

\author{J. Xiang }
\affiliation{School of Nuclear Science and Technology, Lanzhou University, Lanzhou 730000, China}
\affiliation{School of Physical Science and Technology, Southwest University, Chongqing 400715, China}
\author{Z. P. Li }
\affiliation{School of Physical Science and Technology, Southwest University, Chongqing 400715, China}
\author{J. M. Yao }
\affiliation{School of Physical Science and Technology, Southwest University, Chongqing 400715, China}
\affiliation{Department of Physics, Tohoku University, Sendai 980-8578, Japan}
\author{W. H. Long }
\affiliation{School of Nuclear Science and Technology, Lanzhou University, Lanzhou 730000, China}
\author{P. Ring}
\affiliation{Physik-Department der Technischen Universit\"at M\"unchen, D-85748 Garching, Germany}
\affiliation{State Key Laboratory of Nuclear Physics and Technology, School of Physics, Peking University, Beijing 100871, China}
\author{J. Meng }
\affiliation{State Key Laboratory of Nuclear Physics and Technology, School of Physics,
Peking University, Beijing 100871, China}
\affiliation{School of Physics and Nuclear Energy Engineering, Beihang University, Beijing 100191, China}
\affiliation{Department of Physics, University of Stellenbosch, Stellenbosch, South
Africa}

\begin{abstract}
Low-lying nuclear states of Sm isotopes are studied in the framework of a collective Hamiltonian based on covariant energy density functional theory. Pairing correlation are treated by both BCS and Bogoliubov methods. It is found that the pairing correlations deduced from relativistic Hartree-Bogoliubov (RHB) calculations are generally stronger than those by relativistic mean-field plus BCS (RMF+BCS) with same pairing force.
By simply renormalizing the pairing strength, the diagonal part of the pairing field is changed in such a way that the essential effects of the off-diagonal parts of the pairing field neglected in the RMF+BCS calculations can be recovered, and consequently the low-energy structure is in a good agreement with the predictions of the RHB model.
\end{abstract}


\maketitle

The study of nuclear low-lying states is of great importance to unveil the low-energy structure of atomic nuclei and turns out to be essential to understand the evolution of shell structure and collectivity~\cite{Meng98,Hagen12,Kshetri06}, nuclear shape phase transitions~\cite{Meng05,Casten06,Cejnar10}, shape coexistence~\cite{Heyde11}, the onset of new shell gaps~\cite{Ozawa2000}, the erosion of traditional magic numbers~\cite{Sorlin08}, etc. The understanding and the quantitative description of low-lying states in nuclei necessitate an accurate modeling of the underlying microscopic nucleonic dynamics.

Density functional theory (DFT) is a reliable platform  for studying the complicated nuclear excitation spectra and electromagnetic decay patterns~\cite{BHR.03, JacD.11, Vretenar05, Meng06, Meng2013FrontiersofPhysics55}. Since the DFT scheme breaks essential symmetries of the system, this requires to include the dynamical effects related to the restoration of broken symmetries, as well as the fluctuations in the collective coordinates. In recent years several accurate and efficient models and algorithms, based on microscopic density functionals or effective interactions, have been developed that perform the restoration of symmetries broken by the static nuclear mean field, and take the quadrupole fluctuations into account~\cite{NVR.06a,NVR.06b,BH.08,Yao08,Yao.09,Yao.10,RE.10}. This level of implementation is also referred as the multi-reference (MR)-DFT~\cite{Lacroix09}. Compared with MR-DFT, the model of a  collective Hamiltonian with parameters determined in a microscopic way from self-consistent mean-field calculations turns out to be a powerful tool for the systematical studies of nuclear low-lying states~\cite{PR.04,Nik.09,Nik.11}, with much less numerical demanding. Even for the heavy nuclei full triaxial calculations can be relatively easily carried out with a five-dimension collective Hamiltonian \cite{Li.10}.
It has achieved great success in describing the low-lying states in a wide range of nuclei,
from $A\sim 40$ to superheavy nuclei including spherical, transitional, and deformed ones~\cite{Nik.09,Li.09,Li.09b,Li.10,Li.11,Nik.11,Yao11-lambda,Mei12,Del10}.

For open-shell nuclei, pairing correlations between nucleons have important influence on low-energy nuclear structure~\cite{Dean03}. In the relativistic scheme they could be taken into account using the BCS ansatz~\cite{GRT.90} or full Bogoliubov transformation~\cite{KR.91,Rin.96}. Compared with the simple BCS method, the consideration of pairing correlations through the Bogoliubov transformation is numerically demanding for heavy triaxial deformed nuclei. It has been demonstrated that there is no essential difference between BCS and Bogoliubov methods for the descriptions of the ground-state of stable nuclei~\cite{Ring80}.  Girod  {\it et al.} have compared the results obtained from Hartree-Fock-Bogoliubov (HFB) and Hartree-Fock plus BCS (HF+BCS) calculations, including the potential energy surfaces (PESs), pairing gaps, and pairing energies as functions of the axial deformation~\cite{Girod83}. It has been shown that the PESs given by these two methods are very similar. Moreover, the pairing gaps and energies from the HF+BCS calculations are slightly smaller than those from the HFB calculation. In view of these facts, it is natural to test the validity of the BCS ansatz in describing the the low-energy structure of nuclei, as referred to {the RHB method}. Aiming at this point, the comparisons are performed within the covariant density functional based 5DCH model, specifically between the triaxial deformed RMF+BCS and RHB calculations.
Due to the emergence of an abrupt shape-phase-transition~\cite{Li.09}, the even-even Sm isotopes with $134\leqslant A\leqslant 154$ are taken as the candidates in this study.

Practically nuclear excitations determined by quadrupole vibrational and rotational degrees of freedom can be treated by introducing five collective coordinates, i.e., the quadrupole deformations $(\beta,\gamma)$ and Euler angles ($\Omega=\phi,\theta,\psi$)~\cite{Pro.99}. The quantized 5DCH that describes the nuclear excitations of quadrupole vibration, rotation and their couplings can be written as,
\begin{equation}
\label{hamiltonian-quant}
\hat{H} =
\hat{T}_{\textnormal{vib}}+\hat{T}_{\textnormal{rot}}
              +V_{\textnormal{coll}} \; ,
\end{equation}
where $V_{\textnormal{coll}}$ is the collective potential, and $\hat{T}_{\textnormal{vib}}$ and $\hat{T}_{\textnormal{rot}}$ are respectively the vibrational and rotational kinetic energies,
\begin{align}\label{Vcoll}
 {V}_{\text{coll}}  =& E_{\text{tot}}(\beta,\gamma)  - \Delta V_{\text{vib}}(\beta,\gamma) - \Delta  V_{\text{rot}}(\beta,\gamma),\\
\hat{T}_{\textnormal{vib}} =&-\frac{\hbar^2}{2\sqrt{wr}} \left\{\frac{1}{\beta^4}  \left[\frac{\partial}{\partial\beta}\sqrt{\frac{r}{w}}\beta^4
   B_{\gamma\gamma} \frac{\partial}{\partial\beta}\right.\right.\nonumber\\
& \left.\left.- \frac{\partial}{\partial\beta}\sqrt{\frac{r}{w}}\beta^3 B_{\beta\gamma}\frac{\partial}{\partial\gamma} \right]+\frac{1}{\beta\sin{3\gamma}} \left[
   -\frac{\partial}{\partial\gamma} \right.\right.\label{Tvib}\\
& \left.\left.\sqrt{\frac{r}{w}}\sin{3\gamma}  B_{\beta \gamma}\frac{\partial}{\partial\beta} +\frac{1}{\beta}\frac{\partial}{\partial\gamma} \sqrt{\frac{r}{w}}\sin{3\gamma} B_{\beta \beta}\frac{\partial}{\partial\gamma}
   \right]\right\}\nonumber,\\
\hat{T}_{\textnormal{\textnormal{\textnormal{rot}}}}
=&\frac{1}{2}\sum_{k=1}^3{\frac{\hat{J}^2_k}{\mathcal{I}_k}}.\label{Trot}
\end{align}
In eq. (\ref{Vcoll}), $E_{\textnormal{tot}}(\beta,\gamma)$ is the binding energy determined by the constraint mean-field calculations, and the terms $\Delta V_{\textnormal{vib}}$ and $\Delta V_{\textnormal{rot}}$, calculated in the cranking approximation~\cite{Ring80}, are zero-point-energies (ZPE) of vibrational and rotational motions, respectively. In eq. (\ref{Trot}), $\hat{J}_k$ denotes the components of the angular momentum in the body-fixed frame of the nucleus. Moreover the mass parameters $B_{\beta\beta}$, $B_{\beta\gamma}$, $B_{\gamma\gamma}$ in eq. (\ref{Tvib}), as well as the moments of inertia $\mathcal{I}_k$ in eq. (\ref{Trot}), depend on the quadrupole deformation variables $\beta$ and $\gamma$,
\begin{align}
\mathcal{I}_k =& 4B_k\beta^2\sin^2(\gamma-2k\pi/3),&k=&1, 2, 3,
\end{align}
where $B_k$ represents inertia parameter. In eq. (\ref{Tvib}), the additional quantities $r=B_1B_2B_3$ and $w=B_{\beta\beta}B_{\gamma\gamma}-B_{\beta\gamma}^2 $ define the volume element of the collective space. The corresponding eigenvalue problem is solved {by expanding the eigenfunctions on}
a complete set of basis functions in the collective space of the quadrupole deformations $(\beta, \gamma)$ and Euler angles $(\Omega =\phi, \theta, \psi)$.

The dynamics of the 5DCH is governed by seven functions of the intrinsic deformations $\beta$ and $\gamma$: the collective potential $V_{\rm coll}$, three mass parameters $B_{\beta\beta}$, $B_{\beta\gamma}$, $B_{\gamma\gamma}$, and three moments of inertia $\mathcal{I}_k$. These functions are determined using the cranking approximation formula based on the intrinsic triaxially deformed mean-field states. The diagonalization of the Hamiltonian~(\ref{hamiltonian-quant}) yields the excitation energies and collective wave functions that are used to calculate observables~\cite{Nik.09}.

The fact that, the 5DCH model using the collective inertia parameters calculated based on the cranking approximation can reproduce the structure of the experimental low-lying spectra~\cite{Nik.09} up to an overall renormalization factor, demonstrates such approximation is fair enough for the present study. As it has been shown in Ref.~\cite{LNRVYM12}, this factor takes into account the contributions of the time-odd fields. A microscopic calculation of this factor would go far beyond the scope of the present investigation.

The intrinsic triaxially deformed mean-field states are the solutions of the Dirac (RMF+BCS) or RHB equations. The point-coupling energy functional PC-PK1~\cite{Zhao10} and the separable pairing force~\cite{TMR.09a} are used in the particle-hole and particle-particle channels, respectively. In solving the Dirac and RHB equations, the Dirac spinors are expanded on the three-dimension harmonic oscillator basis with 14 major shells \cite{KR.89,Peng08}. A quadratic constraint on the mass quadrupole moments is carried out to obtain the triaxially deformed mean-field states with
$\beta\in[0.0, 0.8]$ and $\gamma\in[0^\circ, 60^\circ]$,and the step sizes $\Delta\beta=0.05$ and $\Delta\gamma=6^\circ$. More details about the calculations can be found in Refs.~\cite{NRV.10,Xiang12}.

\begin{figure}[htbp]
\includegraphics[width = 8cm]{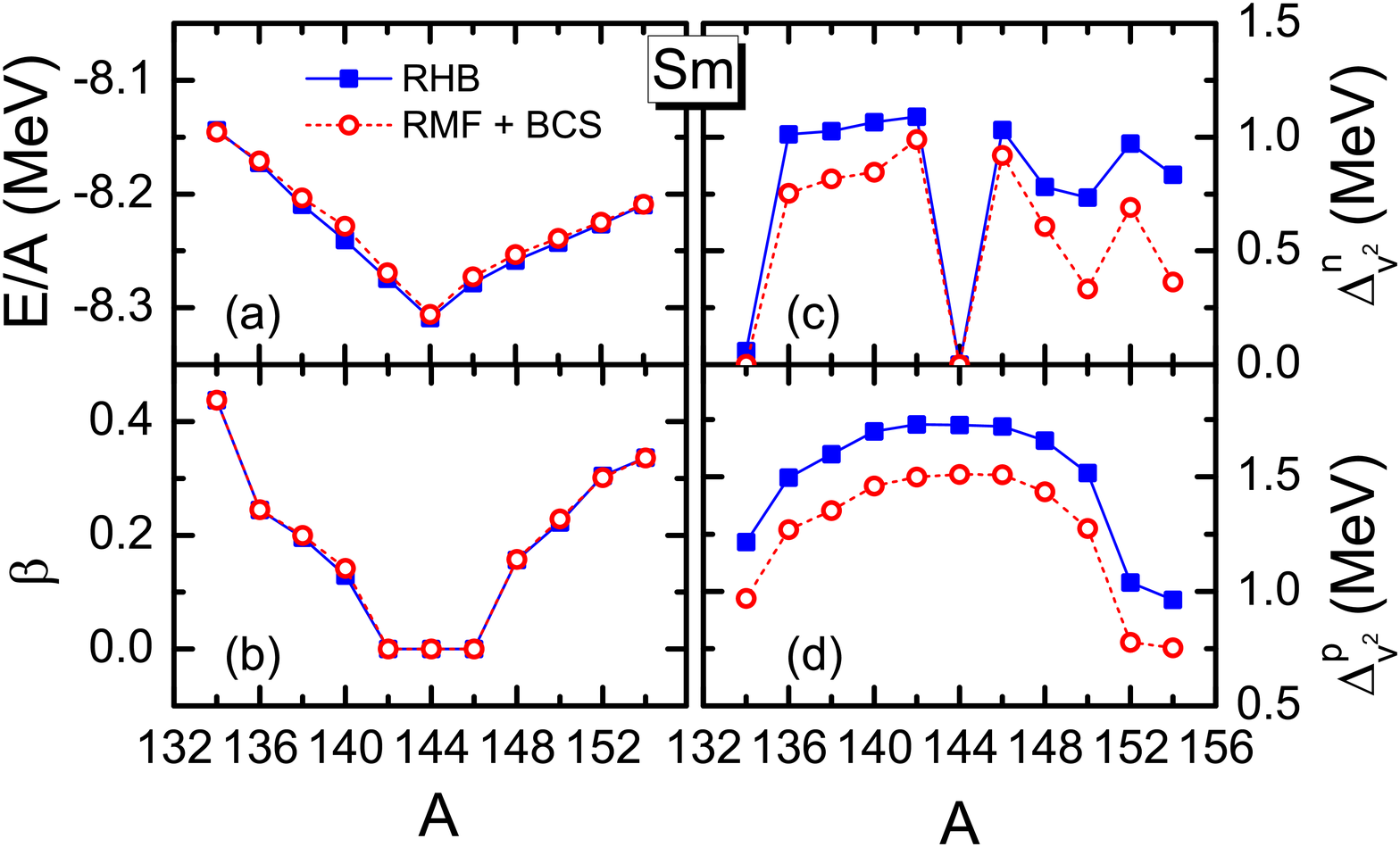}
\caption{\label{fig1}(Color online) Comparison between the RHB and RMF+BCS calculations on the binding energy per nucleon $E/A$ [plot (a)], quadrupole deformation $\beta$ [plot (b)], neutron [plot (c)] and proton [plot (d)] average pairing gaps weighted by the occupation probabilities $v^2$~\cite{Bender00}for even-even Sm isotopes.}
\end{figure}

Figure~\ref{fig1} displays the comparison between the RHB and RMF+BCS calculations for the binding energy per nucleon $E/A$ [plot (a)], quadrupole deformation $\beta$ [plot (b)], neutron [plot (c)] and proton [plot (d)] average pairing gaps weighted by the occupation probabilities $v^2$~\cite{Bender00} of even-even Sm isotopes with $134\leqslant A\leqslant 154$. The binding energies and deformations found in the two calculations are close to each other. However, the average neutron and proton pairing gaps provided by the RHB calculations are generally larger than those by the RMF+BCS ones. This is consistent with the observations in Ref.~\cite{Girod83}, which indicates that the BCS ansatz gives slightly weaker pairing correlations with same pairing force. The underlying reason is well-known that the BCS ansatz corresponds to a special Bogoliubov transformation, which only considers pairing correlation between two nucleons in time-reversed conjugate states~\cite{Ring80}, and the off-diagonal matrix elements of the pairing field $\Delta$ are neglected in this approach.

\begin{figure}[htbp]
\includegraphics[width =8cm]{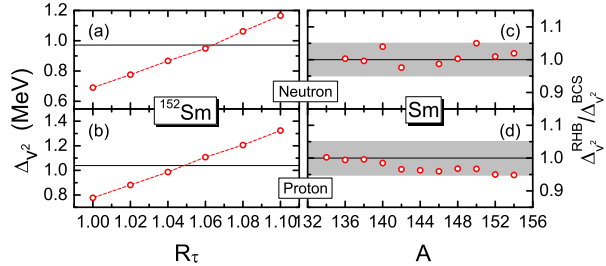}
\caption{\label{fig2}(Color online) Neutron [plot (a)] and proton [plot (b)] average pairing gaps obtained from RMF+BCS calculations as a function of the pairing strength factor $R_\tau$, where the horizontal lines indicate the RHB results with the original pairing force. In the right plots are shown the ratios of the average pairing gaps between the calculations of RHB with the original and RMF+BCS with 6\% enhanced pairing force along the isotopic chain of Sm for neutron [plot (c)] and proton [plot (d)].}
\end{figure}

In the following we have to consider that neglecting the off-diagonal matrix elements of the pairing field leads i) to a reduced configuration mixing and ii) as a consequence of self-consistency also to an overall reduction of the pairing strength in the diagonal matrix elements of the pairing field. Therefore it is interesting to address two points: i) whether the additional configuration mixing induced by the off-diagonal matrix elements of the pairing field is really essential and ii) whether the reduced strength of pairing caused by neglecting the off-diagonal matrix elements in the RMF+BCS approach can recovered simply by multiplying a strength factor $R_\tau$ to the diagonal pairing, i.e. whether the enhanced  pairing strength is also able to reproduce the low-lying structure properties, e.g. the PESs, inertia parameters, as well as the low-lying spectra. Taking $^{152}$Sm as the example, Fig. \ref{fig2} shows the neutron and proton average pairing gaps of the global minimum calculated by RMF+BCS as the functions of the pairing strength factor $R_\tau$, as referred to the horizontal lines denoting the RHB results with original pairing force. It is shown that the average pairing gaps increase almost linearly with respect to the pairing strength factor $R_\tau$ and cross the RHB results at $R_\tau\sim1.06$. Moreover, as shown in Fig. \ref{fig2} (c) and (d) the RMF+BCS calculations with 6\% enhanced pairing strength provide nearly identical average pairing gaps with the RHB results for the selected even-even Sm isotopes, with a relative deviation less than 5\%.

\begin{figure}[htb!]
\includegraphics[width =8cm]{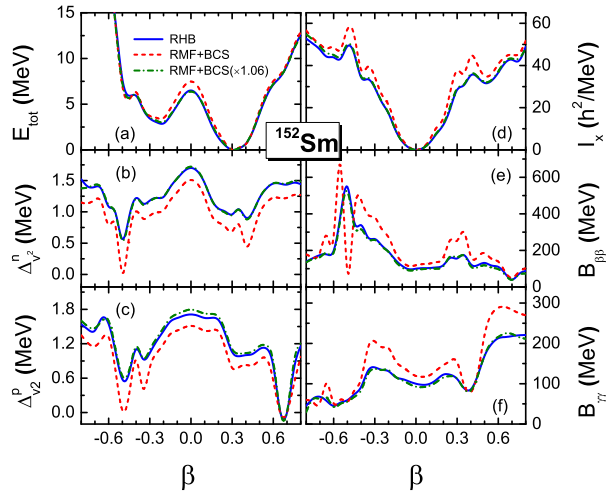}
\caption{\label{fig3}(Color online) Potential energy surfaces (a), neutron (b) and proton (c) average pairing gaps, moments of inertia ${\cal I}_x$ (d), collective masses  $B_{\beta\beta}$ (e) and $B_{\gamma\gamma}$ (f) of $^{152}$Sm as functions of the quadrupole deformation parameter $\beta$ calculated by RHB with the original pairing force (solid lines), and by RMF+BCS with the original (dashed lines) and the enhanced (by 6\%) (dash-dotted lines) pairing force.}
\end{figure}

As the further clarification, Fig. \ref{fig3} displays the PESs, neutron and proton average pairing gaps, moments of inertia ${\cal I}_x$, collective masses  $B_{\beta\beta}$ and $B_{\gamma\gamma}$ for $^{152}$Sm as functions of the quadrupole deformation parameter $\beta$, where the results are calculated by RHB with the original, and by RMF+BCS with the original and the enhanced (by 6\%) pairing strength. It is well demonstrated that for the selected Sm isotopes the deviations on the low-lying structure properties described by RMF+BCS and RHB models can be eliminated by simply enhancing the pairing force about 6\% in the BCS ansatz. Specifically, as the pairing strength increases, the average pairing gaps become larger, which leads to lower spherical barrier of PES~\cite{Rutz99} and reduced inertia parameter~\cite{Sobiczewski69}.

\begin{figure}[htb!]
\includegraphics[width =8cm]{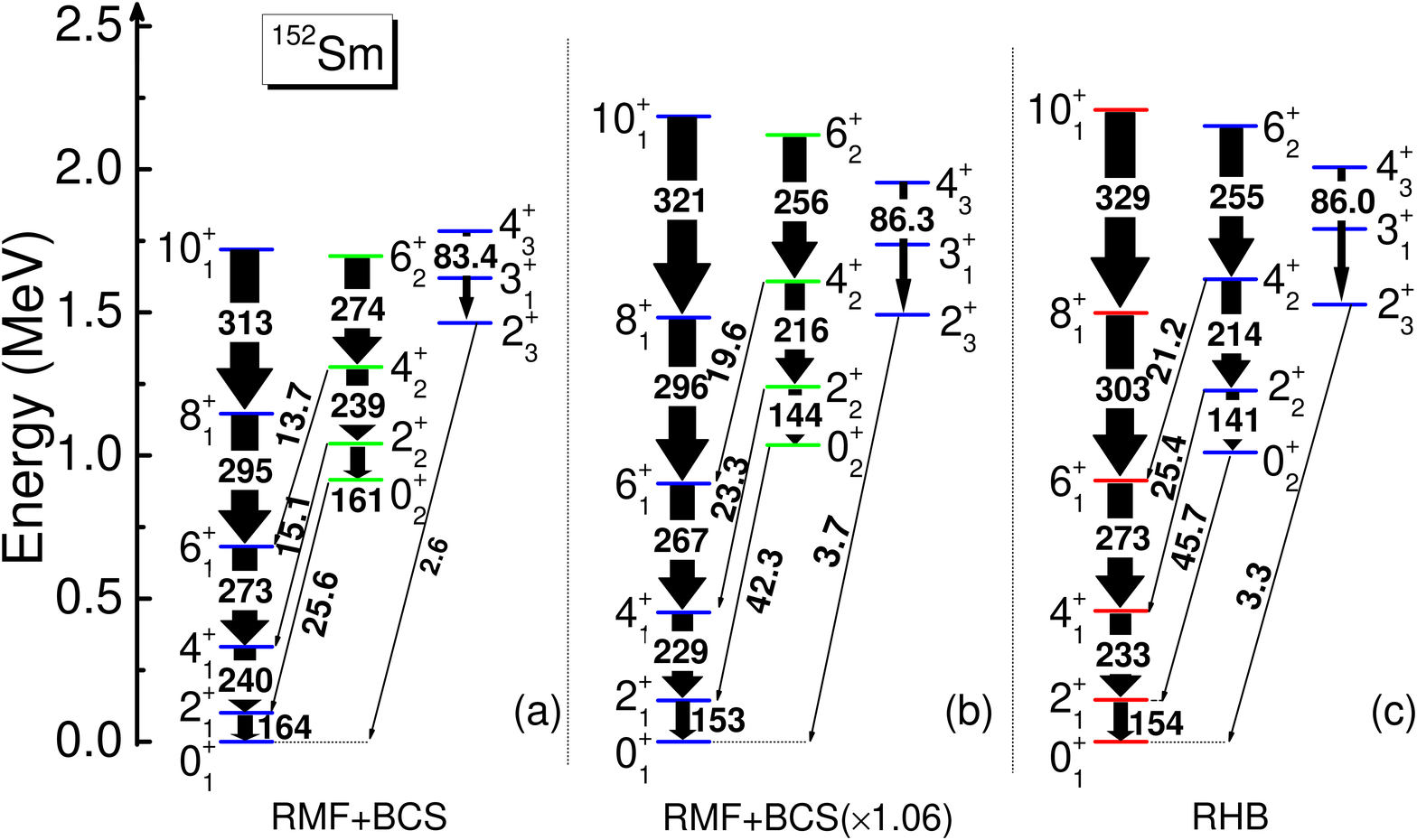}
\caption{\label{fig4}The low-lying spectra of $^{152}$Sm calculated from RMF+BCS with the original [plot a] and the enhanced (by 6\%) [plot b] pairing strength, and compared with results from full RHB calculations [plot c].}
\end{figure}

In Fig.~\ref{fig4} we also compare the theoretical low-lying spectra of $^{152}$Sm calculated by RMF+BCS with the original and the enhanced (by 6\%) pairing strength, to the RHB results. As seen from the left two panels,when the pairing strength is enhanced by 6\%, the low-lying spectrum is extended, and systematically the intraband $B(E2)$ transitions become weaker, and the interband transitions are strengthened, finally leading to an identical prediction as the full RHB calculations (right panel). Quantitatively, the relative deviations between the RHB and RMF+BCS predications are reduced to less than 4\% for the intraband transitions, and the main interband transitions agree with each other within $\sim 2$ W.u.. We have also checked the results for the other Sm isotopes, and very similar spectra are predicted by RHB with the original and RMF+BCS with enhanced (6\%) pairing forces.

\begin{figure}[htb!]
\includegraphics[width =8cm]{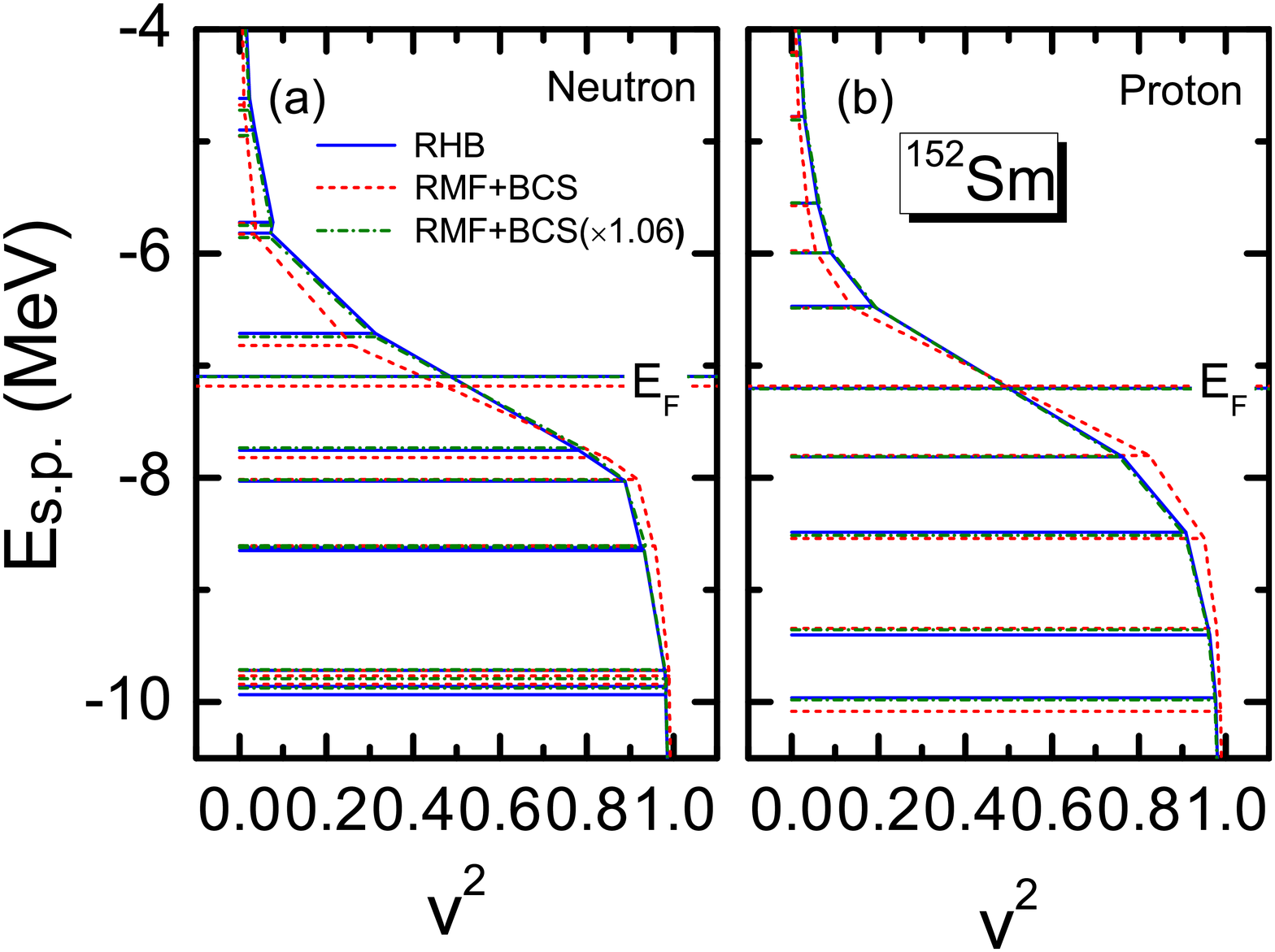}
\caption{\label{fig5}(Color online) Single-particle energy levels (horizontal lines) and occupation probabilities (length of horizontal lines) of $^{152}$Sm calculated by RHB with the original and RMF+BCS with both original and enhanced (by 6\%) pairing strength, where $E_F$ denotes the Fermi levels.}
\end{figure}

The similarity on the low-lying structure can be understood by analyzing the underlying shell structure predicted by the two mean-field calculations. Taking $^{152}$Sm as an example, in Fig.~\ref{fig5} we plot the single-particle configurations (energy and occupation probability) around the Fermi surface corresponding to the mean-field states of the global minimum in the PESs determined by the calculations of RHB with the original and RMF+BCS calculations with both original and enhanced (by 6\%) pairing strength. Notice that the RHB results correspond to the canonical single-particle configurations, which are determined from the diagonalization of the density matrix~\cite{Ring80}. Consistent with the agreement on the low-lying structure properties, the RMF+BCS calculations with the enhanced pairing strength also provide nearly identical single-particle configurations as the RHB ones.

In conclusion, we have taken Sm isotopes as examples to carry out a detailed comparison between the 5DCH calculations based on the RMF+BCS and the RHB approaches for the nuclear low-lying structure properties. It has been shown that the pairing correlations resulting from the RHB method are generally stronger than those from the RMF+BCS method with the same effective pairing force. However, by simply increasing the pairing strength by a factor 1.06 in the RMF+BCS calculations, the low-energy structure becomes very close to that of the full RHB calculations with the original pairing force. We have also carried out similar calculations in other regions of the nuclear chart and found that the necessary renormalization factor stays roughly constant up to heavy nuclei (1.06 in the Pu region) and increases slightly for light ones (1.10 in the Mg region).

This work was supported in part by the Major State 973 Program 2013CB834400, the NSFC under Grant Nos. 11335002, 11075066, 11175002, 11105110, and 11105111, the Research Fund for the Doctoral Program of Higher Education under Grant No. 20110001110087, the Natural Science Foundation of Chongqing cstc2011jjA0376, the Fundamental Research Funds for the Central Universities (XDJK2010B007, XDJK2011B002, and lzujbky-2012-k07), the Program for New Century Excellent Talents in University of China under Grant No. NCET-10-0466, and the DFG cluster of excellence \textquotedblleft Origin and Structure of the Universe\textquotedblright\ (www.universe-cluster.de).


\end{document}